\documentclass[preprint,showpacs,preprintnumbers,amsmath,amssymb]{revtex4}
\usepackage{amsmath}

\usepackage{graphicx}
\usepackage{dcolumn}
\usepackage{bm}
\usepackage{amsmath}

\newcommand{\qed}{\nobreak \ifvmode \relax \else
      \ifdim\lastskip<1.5em \hskip-\lastskip
      \hskip1.5em plus0em minus0.5em \fi \nobreak
      \vrule height0.75em width0.5em depth0.25em\fi}

\begin{document}

\preprint{}

\title{Formulas for Rational-Valued Separability Probabilities of Random Induced Generalized Two-Qubit States}
\author{Paul B. Slater}
 \email{slater@kitp.ucsb.edu}
\affiliation{%
University of California, Santa Barbara, CA 93106-4030\\
}
\author{Charles F. Dunkl}
 \email{cfd5z@virginia.edu}
\affiliation{Department of Mathematics, University of Virginia,
Charlottesville, VA 22904-4137}%

\begin{abstract}
Previously, a formula, incorporating a  $5F4$ hypergeometric function, for the 
 Hilbert-Schmidt-averaged determinantal moments $\left\langle \left\vert \rho^{PT}\right\vert ^{n}\left\vert \rho\right\vert
^{k}\right\rangle /\left\langle \left\vert \rho\right\vert ^{k}\right\rangle$ of $4 \times 4$ density-matrices ($\rho$), and their partial transposes ($\rho^{PT}|$) was applied with $k=0$ to the generalized 
two-qubit separability-probability question. The formula can, further, be viewed we note here,  as an averaging over ``induced measures in the space of mixed quantum states''. The  associated induced-measure  separability probabilities ($k =1, 2,\ldots$) are found--{\it via} a high-precision density approximation 
procedure--to assume interesting, relatively simple rational values in the two-re[al]bit ($\alpha = \frac{1}{2}$), (standard) two-qubit ($\alpha = 1$) and  two-quater[nionic]bit ($\alpha =2$) cases. We deduce rather simple companion (rebit, qubit, quaterbit,\ldots) formulas that successfully reproduce the rational values assumed for {\it general} $k$. These formulas are observed to share certain features, possibly allowing them to be incorporated into a single master formula.
\end{abstract}


\maketitle
\section{Introduction}
The question of the probability that a generic quantum system is separable/disentangled was raised in a 1998 paper of {\.Z}yczkowski, Sanpera, Lewenstein and Horodecki, entitled 
"Volume of the set of separable states" \cite{ZHSL}. Certainly, any particular answer to this question will crucially depend upon the measure that is attached to the systems in question. A large body of literature has arisen from the 1998  study, and we seek to make a significant contribution to it, addressing heretofore unsolved problems. Let us point out the work of Aubrun, Szarek and Ye
\cite{aubrun2}, which addresses questions of a somewhat similar nature to those examined below, while employing the same class of measures. However, their work is set in an {\it asymptotic} framework, while we will be  concerned with obtaining exact {\it finite-dimensional} results (cf. \cite{Bhosale}). On the other hand, Singh, Kunjwal and Simon \cite{Simon} did focus on finite-dimensional scenarios, but with a distinct form of measure, the one originally used in \cite{ZHSL}.

We have investigated the possibility of extending to the class of "induced measures in the space of mixed quantum states" \cite{Induced,ingemarkarol} the line of analysis reported in \cite{MomentBased,slaterJModPhys}, the principal separability probability  findings of which--most notably the two-qubit conjecture of $\frac{8}{33} \approx 0.242424$--have recently been robustly supported, with the use of extensive Monte-Carlo sampling, by Fei and Joynt \cite{FeiJoynt}, as well as by Milz and Strunz, to somewhat similar effect \cite[Fig. 4, eqs. (30), (31)]{milzstrunz} (cf. \cite[Table 1]{Dubna}). This earlier line of work pertained to the use of the Hilbert-Schmidt measure (the particular symmetric case, $K=N$,  of the induced measures) on the high-dimensional convex sets of generalized 
(real-, complex-, quaternionic-entried) two-qubit $(N=4$) states.

In \cite[p. 30]{MomentBased}, a central role had been played by the (not yet formally proven) determinantal moment formula obtained there 
\begin{align*} \label{MomentFormula}
&  \left\langle \left\vert \rho^{PT}\right\vert ^{n}\left\vert \rho\right\vert
^{k}\right\rangle /\left\langle \left\vert \rho\right\vert ^{k}\right\rangle
\\
&  =\frac{\left(  k+1\right)  _{n}\left(  k+1+\alpha\right)  _{n}\left(
k+1+2\alpha\right)  _{n}}{2^{6n}\left(  k+3\alpha+\frac{3}{2}\right)
_{n}\left(  2k+6\alpha+\frac{5}{2}\right)  _{2n}}\\
&  \cdot~_{5}F_{4}\left(
\genfrac{}{}{0pt}{}{-n,-k,\alpha,\alpha+\frac{1}{2},-2k-2n-1-5\alpha
}{-k-n-\alpha,-k-n-2\alpha,-\frac{k+n}{2},-\frac{k+n-1}{2}}%
;1\right)  
\end{align*}
on the basis of extensive computations.
Here $\rho^{PT}$ denotes the partial transpose \cite{asher} of the density matrix $\rho$,  and  $|\rho|$, its determinant, and generalized hypergeometric function notation is employed. The brackets represent averaging with respect to Hilbert-Schmidt measure \cite{szHS}.  Furthermore, $\alpha$ is a random-matrix Dyson-index-like parameter \cite{MatrixModels}, assuming, in particular, the value 1 for the standard (fifteen-dimensional convex set of) density matrices with complex-valued off-diagonal entries.

It subsequently occurred to us that this hypergeometric-based moment formula  
could be readily adapted to the broader class of 
random induced measures by considering, in the notation of 
\cite{Induced,ingemarkarol} that 
\begin{equation}
k = K-N, 
\end{equation}
where $K$ is the dimension of the ancilla/environment state, over which the tracing operation is performed.

As in the earlier work  \cite{MomentBased,slaterJModPhys}, a high-precision density-approximation (inverse) procedure of Provost, 
incorporating the first 11,401 such determinantal moments, strongly indicates that the  random induced-measure  separability 
probabilities ($k =1, 2,\ldots$) assume interesting, relatively simple rational values in the two-re[al]bit ($\alpha = \frac{1}{2}$), (standard) two-qubit ($\alpha = 1$) and  two-quater[nionic]bit ($\alpha =2$) cases, particularly so for $\alpha =1$ (sec.~\ref{Analysis}). One striking example is that for $k=3$, the $\alpha =1$ separability probability is found to be $\frac{27}{38}= \frac{3^3}{2 \cdot 19}$ (to {\it fifteen} decimal places). In fact, based on extensive calculations ($k =0,\ldots,15,\ldots$) of this nature, we are able to deduce rather simple companion (rebit, qubit, quaterbit) formulas 
(\ref{ComplexRule})-(\ref{RebitRule}) that successfully reproduce the rational values assumed for {\it general} integer and half-integer $k$ (sec.~\ref{Companion}).

Further efforts along these lines have been given in a subsequent  paper \cite{HyperDiff}, in which the determinantal inequality $|\rho^{PT}| >|\rho|$ is now imposed, rather than the broader inequality $|\rho^{PT}| >0$.  (Of course, $|\rho| \geq 0$, while $|\rho^{PT}|$ is both a necessary and sufficient condition for separability here \cite{asher,michal}.) There, equivalent hypergeometric- and difference-equation-based formulas, $Q(k,\alpha)= G_1^k(\alpha) G_2^k(\alpha)$, for $k = -1, 0, 1,\ldots,9$, were given for that (rational-valued) portion of the total separability probability satisfying the stricter inequality. (We also preliminarily investigate this problem below in sec.~\ref{Division}.)

Milz and Strunz \cite{milzstrunz} have recently reported a highly interesting finding that the conjectured Hilbert-Schmidt separability probability of $\frac{8}{33} \approx 0.242424$ holds constant along the radius of the Bloch sphere of either of the reduced subsystems of 
generic two-qubit ($\alpha =1$) systems. We are presently investigating the nature that the separability  probability takes as a {\it joint} function of the radii of the two single-qubit subsystems, and related questions.
\section{Analysis} \label{Analysis}
We pursue the indicated extension of our earlier (Hilbert-Schmidt-based) work to random induced measures, in general. As in \cite{MomentBased,slaterJModPhys}, the  determinantal moment formula above is employed in the Legendre-polynomial-based (Mathematica-implemented) density approximation (inverse) procedure of Provost \cite{Provost}. This  possesses a least-squares rationale. The program as originally presented is speeded, by incorporating the well-known recursion formula for Legendre polynomials, so that successive  polynomials do not have to be computed {\it ab initio}. The computations are all exact, in nature, rather than numerical. Provost advises that the procedure should be 
regarded as an "approximation", rather than an "estimation" scheme \cite{Provost}. (Let us note that the implementation of the procedure requires considerable caution and an adaptive strategy when the term 
$(k-j+1)_{n-j}$ \cite[sec. D.2]{MomentBased} in the underlying summation formula for the hypergeometric-based determinantal 
moments is zero. It is zero if $k-j+1 \leq 0 \leq k+n-2 j$, that is, if values $j$  for which $k+1 \leq j$ and $2 j \leq k+n$ occur in the summation $j=0...n$.)

Now, with the use of an unprecedentedly large 
number (11,401) of the determinantal moments,
we found (to ten decimal places) for $k =1$, the separability probability
of the standard, complex ($\alpha =1$) 15-dimensional convex set of two-qubit states to be 
$\frac{61}{143} =\frac{61}{11 \cdot 13} \approx 0.4265734$. On the other hand,
for the Hilbert-Schmidt case ($k = K - N = 0$), a very compelling body of evidence of a number of types 
(though yet no formal proof) has been adduced that the corresponding separability
probability, as has been already noted, is $\frac{8}{33} =\frac{2^3}{3 \cdot 11} \approx 0.242424$ \cite{MomentBased,slaterJModPhys,FeiJoynt,milzstrunz}.

For the quaternionic ($\alpha =2$) case, the induced-measure ($k=1$) separability
probability (now to thirteen decimal places) was $\frac{3736}{22287} =  
\frac{2^3 \cdot 467}{3 \cdot 17 \cdot 19 \cdot 23} \approx 0.16763135$, while the Hilbert-Schmidt counterpart strongly appears to be $\frac{26}{323} = 
\frac{2 \cdot 13}{17 \cdot 19} \approx 0.0804953$ \cite{MomentBased,slaterJModPhys,FeiJoynt}. 

Let us further note, though any immediate
quantum-mechanical random-matrix division-algebra interpretation does not seem at hand, that for 
$k = 1, \alpha = 3$, we obtain a "separability-probability" approximant, based on the 11,401 moments, that, to a remarkable 
{\it sixteen}
decimal places equalled 
$\frac{8159}{124062}  = \frac{41 \cdot 199}{2 \cdot 3 \cdot 23 \cdot  29 \cdot 31} \approx 
0.0657655$. This particularly high accuracy appears to essentially be an artifact
of the Legendre-polynomial-based procedure that commences with a {\it uniform}
distribution over the the interval $|\rho| \in [-\frac{1}{16},\frac{1}{256}]$. For such a distribution, 
the probability over the "separability" interval of $[0, \frac{1}{256}]$ is 
the ratio of $\frac{1}{256}$ to $(\frac{1}{16} +\frac{1}{256})$, that is  $\frac{1}{17} 
\approx 0.0588235$, quite near to 0.0657655. So as separability probability approximants  increasingly deviate from the uniform-based  one of $\frac{1}{17}$, at least for specific $k$, we can expect convergence of the density-approximation
procedure to relatively weaken.

For the two-rebit scenario ($\alpha = \frac{1}{2}$), the associated Hilbert-Schmidt
separability probability strongly  appears to be $\frac{29}{64} = \frac{29}{2^6} \approx 
0.453125$ \cite{MomentBased,slaterJModPhys}, while in the random induced-measure $k = 1$ counterpart, we obtain (to almost nine  decimal places) a value once again larger than that for the Hilbert-Schmidt case of $k=0$, that is, $\frac{515}{768} =\frac{5 \cdot 103}{2^8 \cdot 3} \approx 0.670573$. (Note the powers of 2, in both denominators--a phenomenon 
that will continue to be observed for rebit-related results.)

\begin{table}
\label{table:TwoRebits}
\caption{Two-Rebit ($\alpha = \frac{1}{2}$) Separability Probabilities}
\centering
  \begin{tabular}{c || c | c | c}
    \hline\hline
    $k = 0$ & $\frac{29}{64}$  & $\frac{29}{2^6}$  & 0.453125 \\ \hline
        $k = 1$ & $\frac{515}{768}$ &  $\frac{5 \cdot 103}{2^8 \cdot 3}$ &  0.670573\\
    \hline
$k = 2$ & $\frac{1645}{2048}$ & $\frac{5 \cdot 7 \cdot 47}{2^{11}}$ & 0.803222 \\  \hline
$k =3 $ &$\frac{31641}{35840}$ & $\frac{3 \cdot 53 \cdot 199}{2^{10} \cdot 5 \cdot 7}$ & 
0.882840\\ \hline
$k =4 $ &$\frac{274373}{294912}$ & $\frac{11 \cdot 24943}{2^{15} \cdot 3^2}$ & 
0.930355 \\ \hline
$k =5 $ &$\frac{439777}{458752}$ & $\frac{13 \cdot 33829}{2^{16} \cdot 7}$ & 
0.958638 \\ \hline
$k =6 $ &$ \frac{11251151}{11534336}$ & $\frac{11251151}{2^{20} \cdot 11}$ & 
0.975448 \\ \hline
$k =7 $ &$ \frac{30224045}{30670848}$ & $\frac{5 \cdot 17 \cdot 53 \cdot 6709}{2^{18} \cdot 3^2 \cdot 13}$ & 
0.985432 \\ \hline
$k =8 $ &$ \frac{10395147}{10485760}$ & $\frac{3 \cdot 7 \cdot 19 \cdot 26053}{2^{21} \cdot 5}$ & 
0.991358 \\ \hline
\hline
  \end{tabular}
\end{table}
\begin{table}
\caption{Two-Qubit ($\alpha = 1$) Separability Probabilities}
\centering
  \begin{tabular}{c || c | c | c}
    \hline\hline
    $k = 0$ & $\frac{8}{33}$  & $\frac{2^3}{3 \cdot 11}$  & 0.242424 \\ \hline
    $k = 1$ & $\frac{61}{143}$ & $\frac{61}{11 \cdot 13}$ &   0.426573\\
    \hline
$k = 2$ & $\frac{259}{442}$ & $\frac{7 \cdot 37}{2 \cdot 13 \cdot 17}$ & 0.585973 \\  \hline
$k =3 $ &$\frac{27}{38}$ & $\frac{3^3}{2 \cdot 19}$ & 
0.710526 \\ \hline
$k = 4$ & $\frac{5960}{7429}$ & $\frac{2^3 \cdot 5 \cdot 149}{17 \cdot 19 \cdot 23}$ & 
0.802261 \\ \hline
$k = 5$ & $\frac{379}{437}$ & $\frac{379}{19 \cdot 23}$ & 
0.867277 \\
\hline
$k = 6$ & $\frac{63881}{70035}$ & $\frac{127 \cdot 503}{3 \cdot 5 \cdot 7 \cdot 23 \cdot 29}$ & 
0.912129\\ \hline 
$k = 7$ & $\frac{1169237}{1240620}$ & $\frac{37 \cdot 31601}{2^2 \cdot 3 \cdot 5 \cdot 23 \cdot 29 \cdot 31}$ & 
0.942461\\ \hline 
$k = 8$ & $\frac{25963}{26970}$ & $\frac{7 \cdot 3709}{2 \cdot 3 \cdot 5 \cdot 29  \cdot 31}$ & 
0.962662\\
\hline
  \end{tabular}
\label{table:TwoQubits}
\end{table}
\begin{table}
\caption{Two-Quaterbit ($\alpha =2$) Separability Probabilities}
\centering
  \begin{tabular}{c || c | c | c}
    \hline\hline
    $k = 0$ & $\frac{26}{323}$  & $\frac{2 \cdot 13}{17 \cdot 19}$  & 0.080495 \\ \hline
    $k = 1$ & $\frac{3736}{22287}$ &  
$\frac{2^3 \cdot 467}{3 \cdot 17 \cdot 19 \cdot 23} $ & 0.167631\\
    \hline
$k = 2$ & $\frac{1807}{6555} $ & $\frac{13 \cdot 139}{3 \cdot 5 \cdot 19 \cdot 23}$ & 0.275667 \\  \hline
$k =3 $ &$\frac{3919}{10005}$ & $\frac{3919}{3 \cdot 5 \cdot 23 \cdot 29}$ & 
0.391704\\ \hline
$k =4 $ &$\frac{104379}{206770}$ & $\frac{3 \cdot 11 \cdot 3163}{2 \cdot 5 \cdot 23 \cdot 29 \cdot 31}$ & 
0.504807 \\ \hline
$k =5 $ &$\frac{16387}{26970}$ & $\frac{7 \cdot 2341}{2 \cdot 3 \cdot 5 \cdot 29 \cdot 31}$ & 
0.607601 \\ \hline
$k =6 $ &$\frac{69475}{99789}$ & $\frac{5^2 \cdot 7 \cdot 397}{3 \cdot 29 \cdot 31 \cdot 37}$ & 
0.696219 \\ \hline
$k =7 $ &$\frac{203123}{263958}$ & $\frac{229 \cdot 887}{2 \cdot 3 \cdot 29 \cdot 37 \cdot 41}$ & 
0.769527 \\ \hline
$k =8 $ &$\frac{1674746}{2022161}$ & $\frac{2 \cdot 837373}{31 \cdot 37 \cdot 41 \cdot 43}$ & 
0.828196 \\ \hline
\hline
  \end{tabular}
\label{table:TwoQuaterbits}
\end{table}

In Tables I, II and III, we present our conclusions, based on such high-precision  calculations, as to the 
rational values ($k=0,1, \ldots 8$) assumed by these induced-measure separability probabilities. Let us note that with the sole exception of $k=7$, the rational values assumed by the (standard) two-qubit ($\alpha =1$) induced states have both smaller denominators and numerators than the other two cases tabulated, indicative, presumably, 
in some manner, of their greater "naturalness".
\section{Three Companion Separability Probability Formulas} \label{Companion}
Further extending the entries of the two-qubit table (Table II), but not explicitly showing the results here, to $k=17$, application of the Mathematica command FindSequenceFunction  to the sequence of length eighteen obtained, plus subsequent simplification procedures, yielded the  governing rule
\begin{equation} \label{ComplexRule}
P^{qubit}_k=1-\frac{3\ 4^{k+3} (2 k (k+7)+25) \Gamma \left(k+\frac{7}{2}\right) \Gamma (2k+9)}{\sqrt{\pi } \Gamma (3 k+13)}.
\end{equation}
Here $P^{qubit}_k$ is the separability probability of the (15-dimensional) standard, complex two-qubit systems endowed with the induced measure $k =K-4$. This formula, thus, successfully reproduces the entries of Table II, as well as the subsequent ones ($k=9,\ldots,17$) we have approximated to high precision, making use of the 11,401 moments in the Provost Legendre-polynomial-based algorithm. (For $k=0$, formula (\ref{ComplexRule}) does, in fact, yield the apparent Hilbert-Schmidt separability probability of $\frac{8}{33}$ \cite{MomentBased,slaterJModPhys,FeiJoynt} [Table II].)

Similarly, employing a somewhat longer sequence $k = 0,\ldots,21$, we obtained the quaternionic ($\alpha =2$) counterpart 
\begin{equation} \label{QuaternionicRule}
P^{quaterbit}_k=1-\frac{4^{k+6} (k (k (2 k (k+21)+355)+1452)+2430) \Gamma
   \left(k+\frac{13}{2}\right) \Gamma (2 k+15)}{3 \sqrt{\pi } \Gamma (3 k+22)},
\end{equation}
yielding the $k=0$ (Hilbert-Schmidt) value of $\frac{26}{323}$.
Further, for the rebit ($\alpha = \frac{1}{2}$) scenario, making analogous use of the sequence 
$k =0,\ldots,15$, we found 
\begin{equation} \label{RebitRule}
P^{rebit}_k=1-\frac{4^{k+1} (8 k+15) \Gamma (k+2) \Gamma \left(2
   k+\frac{9}{2}\right)}{\sqrt{\pi } \Gamma (3 k+7)},
\end{equation}
yielding for $k=0$, the result $\frac{29}{64}$.

\begin{figure}
\includegraphics{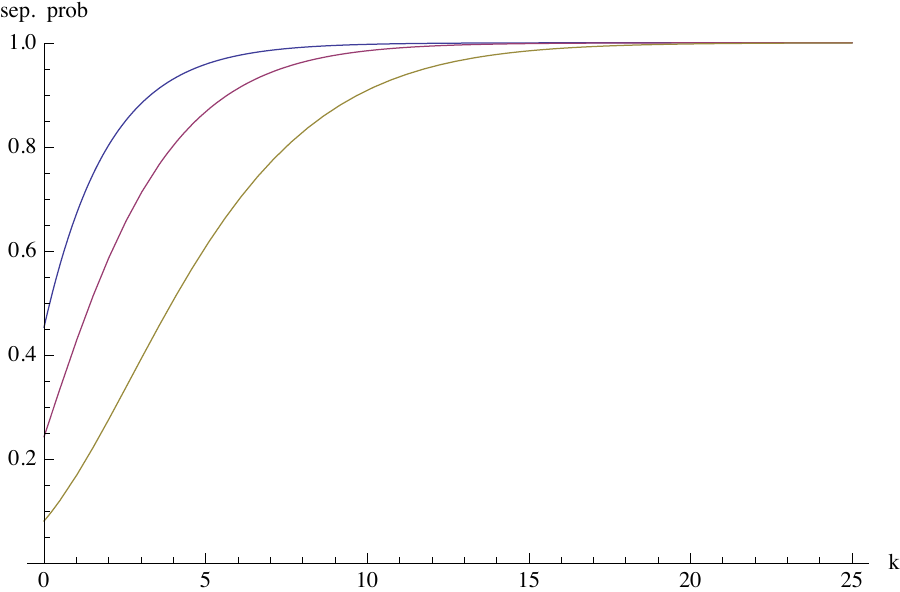}
\caption{\label{fig:Formulas}Two-rebit $>$ two-qubit $>$ two-quaterbit separability probabilities--given by (\ref{RebitRule}), (\ref{ComplexRule}) and (\ref{QuaternionicRule}), respectively--as functions of $k=K-4$}
\end{figure}

In Fig.~\ref{fig:Formulas} we show a joint plot of these three separability probability formulas, with  the rebit one ($\alpha = \frac{1}{2}$) dominating the qubit one ($\alpha =1$), which in turn dominates the quaterbit ($\alpha =2$) curve.
In the limit $k \rightarrow \infty$, the three curves/probabilities  all approach 1 
(cf. \cite{aubrun2}). We have found \cite[sec. III]{HyperDiff}--through analytic means--that for each of $\alpha = 1,2,3,4$ and $\frac{1}{2}, \frac{3}{2},\frac{5}{2},\frac{9}{2}$, that as $k \rightarrow \infty$, the ratio of the logarithm of the 
$(k+1)$-st separability probability to the logarithm of the 
$k$-th separability probability is $\frac{16}{27}$. (Presumably, the pattern continues for larger $\alpha$, but the required computations have, so far, proved too challenging.)

It is interesting to observe, additionally, that for $k=-1$ (that is, $K=3$), a value not apparently susceptible to use of the principal $5F4$-hypergeometric determinantal moment formula and the density approximation (inverse) procedure of Provost \cite{Provost}, the three basic formulas yield
the (now {\it smaller} than Hilbert-Schmidt) further simple rational values $\frac{1}{8}, \frac{1}{14}$ and $\frac{11}{442}$, for the rebit, qubit and quaterbit cases, respectively 
(cf. \cite[p. 130]{aubrun2}). Further, for $k=-2$ $(K = 2)$, the rebit formula has a singularity, the qubit formula yields 0, and the quaterbit one gives $\frac{1}{429} = \frac{1}{3 \times 11 \times 13} \approx 0.002331$.

We have been able to formally extend this series of three formulas to other values of 
$\alpha$, as well, including $\alpha =\frac{3}{2}, \frac{5}{2}, 3,\frac{7}{2}, 4 ,
\frac{9}{2}, 5, 6,\ldots,13$ obtaining similarly structured (increasingly larger) formulas. A major challenge that we are continuing to address is to find a {\it single master} formula
that encompasses these several results, and can itself yield the formula for any specific half-integer or integer value of $\alpha$ (Appendix I).

\section{Division of Separability Probabilities Based on Determinantal Inequalities} \label{Division}
We have also begun to investigate related aspects of the geometry of random-induced generalized two-qubit states, making use of a {\it second}  hypergeometric-based determinantal moment formula
\cite[sec. II]{WholeHalf}
\begin{align*}
\left\langle \left\vert \rho\right\vert
^{k}\left(  \left\vert \rho^{PT}\right\vert -\left\vert \rho\right\vert
\right)  ^{n}\right\rangle /\left\langle \left\vert \rho\right\vert
^{k}\right\rangle   &  =
\left(  -1\right)  ^{n}\frac{\left(  \alpha\right)  _{n}\left(  \alpha
+\frac{1}{2}\right)  _{n}\left(  n+2k+2+5\alpha\right)  _{n}}{2^{4n}\left(
k+3\alpha+\frac{3}{2}\right)  _{n}\left(  2k+6\alpha+\frac{5}{2}\right)
_{2n}}\\
& \times~_{4}F_{3}\left(
\genfrac{}{}{0pt}{}{-\frac{n}{2},\frac{1-n}{2},k+1+\alpha,k+1+2\alpha
}{1-n-\alpha,\frac{1}{2}-n-\alpha,n+2k+2+5\alpha}%
;1\right)  .
\end{align*}
The range of the determinant difference variable $(|\rho^{PT}|-|\rho|)$ is $[-\frac{1}{16},\frac{1}{432}]$, and we shall approximate the contributions over $[0,\frac{1}{432}]$ to the total separability probabilities given in Tables I, II and III.

In \cite{WholeHalf}, employing the first 9,451 of these moments (having set $k$ to zero) in the density approximation procedure of Provost \cite{Provost}, we obtained highly  convincing numerical evidence that the basic set of three Hilbert-Schmidt separability
probabilities ($\frac{29}{64}, \frac{8}{33}, \frac{26}{323}$) was evenly (symmetrically) split
(that is, $\frac{29}{128}, \frac{4}{33}, \frac{13}{323}$) between the two scenarios
$|\rho^{PT}| >|\rho|$ and $|\rho| > |\rho^{PT}| >0$.
Now, with the use of 14,051 such determinantal moments,
with $k=1$, $\alpha =1$, we obtained an approximant equal to eight decimal places to 
$\frac{45}{286}= \frac{3^2 \cdot 5}{2 \cdot 11 \cdot 13} \approx 0.157342657$
for the case 
$|\rho^{PT}| >|\rho|$. Employing the total separability probability $k=1$ result of 
$\frac{61}{143}$ in Table II, we find a  complementary (larger) approximant of   $\frac{7}{26}= \frac{7}{2 \cdot 13} \approx 0.269230769$. So, the symmetry present in the Hilbert-Schmidt case (for example, $\frac{8}{33}= \frac{4}{33}+\frac{4}{33}$) is lost for $k \neq 0$. 

Similarly, 
for the $k=1$, $\alpha =2$ counterpart, we obtain an approximant equal, to almost twelve decimal places,  to  $\frac{2056}{37145} =\frac{2^3 \cdot 257}{5 \cdot 17 \cdot 19 \cdot 23} \approx 0.0553506528470$, when $|\rho^{PT}| > |\rho|$, and, thus, 
$\frac{32}{285} =\frac{2^5}{3 \cdot 5 \cdot 19} \approx 0.1122807017544$ for the complementary 
(larger) approximant. 

To complete the basic triad, that is 
$k=1$ and $\alpha =\frac{1}{2}$ (for which, convergence is typically weakest), for  
$|\rho^{PT}| > |\rho|$, we have an approximant equal, to more than six decimal places, to 
$\frac{281}{1024}=\frac{281}{2^{10}} \approx 0.2744140625$,  and a complementary (larger, again) approximant of  $\frac{1217}{3072}=\frac{1217}{2^{10} \cdot 3} \approx 0.3961588542$. (Note, again, the occurrence of powers of 2 in the $\alpha=\frac{1}{2}$ case.)

For $k = -1$, $\alpha =2$ it is interesting to note that the approximation of the probability
$|\rho^{PT}| > |\rho|$ is $\frac{11}{442}$ to ten decimal places. This is the {\it same}
rational value we found above for the {\it total} separability probability. It, thus, appears that we can conclude that the complementary probability (that is, for 
$|\rho| > |\rho^{PT}| >0$) is now {\it smaller}, in fact, zero, in contrast to 
the $k = 1$ case. The complementary probability also appears to be zero for the two companion cases of $k = -1$, $\alpha =1$ and $\alpha =\frac{1}{2}$.

\begin{figure}
\includegraphics{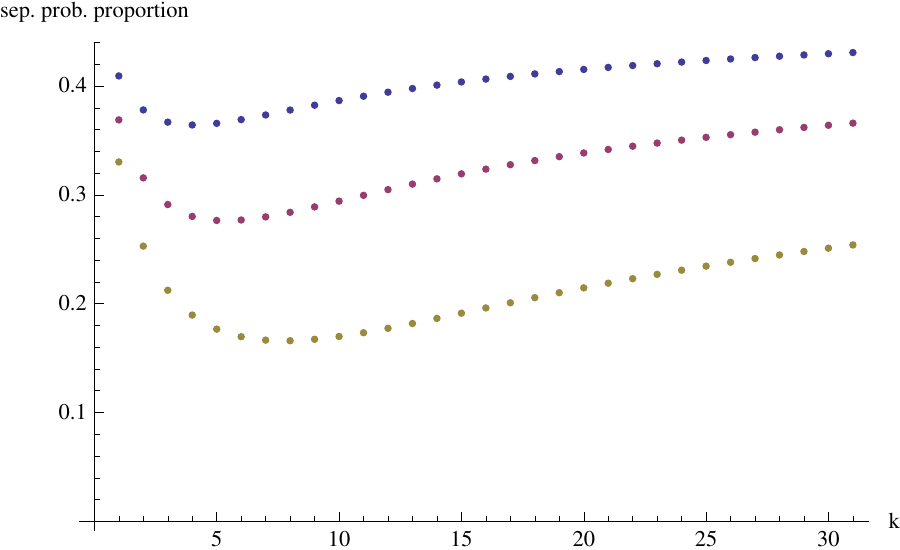}
\caption{Proportion of the total random-induced separability probabilities--based on 9,201 
moments--accounted for by the region $|\rho^{PT}| > |\rho|$. 
The two-rebit ($\alpha =\frac{1}{2}$) curve is dominant, the two-qubit ($\alpha = 1$), intermediate, and the two-quaterbit ($\alpha = 2$) curve, subordinate.}
\label{fig:Proportion}
\end{figure}

In Figure~\ref{fig:Proportion}, we show--based on numerical results using 9,201 moments--the proportion of the three basic total random induced separability probabilities (Tables I, II, III), as a function of $k$,  accounted for by the region $|\rho^{PT}| > |\rho|$.
We have been investigating the possibility of obtaining explicit formulas--as we have been able to do above ((\ref{ComplexRule}),(\ref{QuaternionicRule}),(\ref{RebitRule})) for the total separability probabilities (that is, independently of whether $|\rho|> |\rho^{PT}| >0$ or $|\rho^{PT}| >|\rho|$)--for these sets of complementary probabilities. To even hope to achieve such a goal, it seems necessary to fill in considerably more rows of Table IV 
than we have so far been able to do (cf. \cite{HyperDiff}). 
\section{Alternative Density Approximation Procedure}
In pursuit of such a goal, we have developed an alternative (Appendix II) to the Legendre-polynomial-based density approximation procedure of Provost \cite{Provost}, which we have made abundant use of above and in our earlier work \cite{MomentBased,slaterJModPhys,WholeHalf}. Though well-conditioned, it perhaps is relatively slow to converge for our purposes,
since it takes as the baseline (starting) distribution, the uniform
one, which is far from the sharply-peaked ones, with vanishing endpoints, we have encountered in our separability probability investigations.
The approach outlined in Appendix II uses base functions of the form ${((x-a)(b-x))}^{\alpha}$ where $\alpha$ is a small positive integer.
(Provost does present a number of codes, other than the 
Legendre-polynomial one, including one based on Jacobi polynomials 
\cite[pp. 15, 24]{Provost}. It employs an adaptive strategy of matching the first and second given moments to those of a beta distribution. But we have found this algorithm to be highly 
ill-conditioned in our specific applications.)

\section{Conclusions}
We have reported above some considerable successes in our effort to extend to random induced measures \cite{Induced,ingemarkarol}, earlier separability probability work \cite{MomentBased,slaterJModPhys} based on the Hilbert-Schmidt measure (the particular symmetric $N=K$ case of the random induced measures), and the inequality $|\rho^{PT}|>0$. Further efforts using the more restrictive inequality $|\rho^{PT}] > |\rho|$ utilized in  
sec.~\ref{Division} have been given in a subsequent  paper \cite{HyperDiff}. There equivalences between certain hypergeometric-based formulas and difference equations have been noted.

Let us importantly note that in the recent study \cite{LatestCollaboration2} the (random induced measure) separability probability  problems posed above, have, in fact, been exhaustively {\it formally} solved for the ``toy'' seven-dimensional $X$-states model \cite{Xstates2} of $4 \times 4$ density matrices. Here, contrastingly, we  have concentrated on the more general cases of $4 \times 4$ density matrices  with none of the  off-diagonal entries {\it a priori} nullified (as they are in the $X$-states model). Although, we have developed certain formulas here, for which the evidentiary support is quite considerable, we still lack formal proofs in this higher-dimensional venue. 

We continue to investigate these problems 
in search of a still more definitive ("master formula") resolution of them (Appendix I). As a possible tool in such research, we have developed (Appendix II) an alternative density approximation
procedure to that of Provost \cite{Provost}, on which we have strongly relied to this point in obtaining exact separability probability results.

\begin{table}
\label{table:Proportions}
\caption{Proportions of total separability probabilities $|\rho^{PT}|>|\rho|$}
\centering
  \begin{tabular}{c || c | c | c}
    \hline\hline
 $\alpha $ & $\frac{1}{2}$  & 1 & 2 \\ \hline \hline       
   $k = 0$ & $\frac{1}{2}$  & $\frac{1}{2}$  & $\frac{1}{2}$ \\ \hline
    $k=1$ & $\frac{843}{2060}$ &  $\frac{45}{122}$ & $\frac{771}{2335}$ \\ \hline
$k=2$ & $\frac{9949}{26320}$ &  $\frac{1553}{4921}$ & $\frac{26503}{104806}$ \\ \hline
$k=3$ &---&  $\frac{3073}{10557}$ & $\frac{51585}{242978}$ \\ \hline
$k=4$ &---&  $\frac{2087}{7450}$ & $\frac{2195945}{11586069}$ \\ \hline
$k=5$ &---&---& $\frac{4390079}{24859079}$ \\ \hline
        \hline
$k=6$ &---&  --- & $\frac{8310451}{48993770}$ \\ \hline
        \hline
\end{tabular}
\end{table}
\section{Appendix I. Master Formula Investigation}
This appendix is based on the random induced measure separability probability formulas 
we have obtained for $\alpha=\frac{1}{2},\frac{3}{2},\frac{5}%
{2},\frac{7}{2},1,\ldots13\medskip$.

The purpose is to find $P\left\{  \left\vert \rho^{PT}\right\vert >0\right\}
$ with respect to the normalized measure $\left\vert \rho\right\vert ^{k}$
with parameter $\alpha$. The values $\alpha=\frac{1}{2},1,2$ correspond to the
real, complex, quaternion cases respectively. The obtained formulas have the
form%
\[
P\left\{  \left\vert \rho^{PT}\right\vert >0\right\}  =1-F\left(
\alpha,k\right)  .
\]
Define%
\[
G\left(  \alpha,k\right)  :=4^{k}\frac{\Gamma\left(  k+3\alpha+\frac{3}%
{2}\right)  \Gamma\left(  2k+5\alpha+2\right)  }{\Gamma\left(  \frac{1}%
{2}\right)  \Gamma\left(  3k+10\alpha+2\right)  }.
\]

The first observation: when $\alpha$ is integer or half-integer $\dfrac
{F\left(  \alpha,k\right)  }{G\left(  \alpha,k\right)  }$ is a rational
function of $k$, that is, a ratio of polynomials.

The second observation: when $\alpha$ is an integer then%
\[
F\left(  \alpha,k\right)  =p_{\alpha}\left(  k\right)  G\left(  \alpha
,k\right)  ,
\]
where $p_{\alpha}\left(  k\right)  $ is a polynomial of degree $4\alpha-2$
with leading coefficient $\dfrac{2^{8\alpha+1}}{\left(  2\alpha-1\right)  !},$
and $p_{\alpha}$ can be factored as $\left(  k+g_{1}\left(  \alpha\right)
\right)  \left(  k+g_{1}\left(  \alpha\right)  +1\right)  \cdots\left(
k+g_{2}\left(  \alpha\right)  \right)  \widetilde{p_{\alpha}}\left(  k\right)
$, where $\widetilde{p_{\alpha}}\left(  k\right)  $ is irreducible in general;
furthermore%
\begin{align*}
g_{1}\left(  \alpha\right)   &  :=2\alpha+1+\left\lfloor \frac{\alpha+1}%
{2}\right\rfloor ,\\
g_{2}\left(  \alpha\right)   &  :=3\alpha+\left\lfloor \frac{\alpha+1}%
{3}\right\rfloor .
\end{align*}
The sequence of values $\left[  g_{1}\left(  \alpha\right)  ,g_{2}\left(
\alpha\right)  \right]  $ for $\alpha=2,\ldots,14$ is%
\begin{align*}
&  \left[  6,7\right]
,[9,10],[11,13],[14,17],[16,20],[19,23],[21,27],[24,30],[26,33],[29,37],\\
&  \lbrack31,40],[34,43],[36,47]
\end{align*}

These conjectures imply that the degree of $\widetilde{p_{\alpha}}\left(
k\right)  $ is%
\[
4\alpha-2-\left(  g_{2}\left(  \alpha\right)  +1-g_{1}\left(  \alpha\right)
\right)  =3\alpha+\left\lfloor \frac{\alpha+1}{2}\right\rfloor -\left\lfloor
\frac{\alpha+1}{3}\right\rfloor -2.
\]

The coefficient of $k^{4\alpha-3}$ in $\left(  \frac{2^{8\alpha+1}}{\left(
2\alpha-1\right)  !}\right)  ^{-1}p_{\alpha}\left(  k\right)  $ (note that
this is monic, coefficient of $k^{4\alpha-2}$ is $1$) is given by%
\[
c_{2}\left(  \alpha\right)  :=-3+\frac{3}{2}\alpha+\frac{17}{2}\alpha
^{2}+\left(  \left\lfloor \frac{\alpha-1}{4}\right\rfloor -\left\lfloor
\frac{\alpha}{4}\right\rfloor \right)  \left(  1+\frac{5}{2}\alpha\right)  ,
\]
equivalently%
\[
c_{2}\left(  \alpha\right)  =\left\{
\begin{array}
[c]{c}%
-4-\alpha+\frac{17}{2}\alpha^{2},\alpha\equiv0\operatorname{mod}4,\\
-3+\frac{3}{2}\alpha+\frac{17}{2}\alpha^{2},\alpha\neq0\operatorname{mod}4.
\end{array}
\right.
\]
To determine the second coefficient of $\widetilde{p_{\alpha}}$ note that the
second coefficient of $\left(  k^{n}+a_{2}k^{n-1}+\ldots\right)  \left(
k^{m}+b_{2}k^{m-1}+\ldots\right)  =k^{n+m}+\left(  a_{2}+b_{2}\right)
k^{n+m-1}+\ldots$ is $a_{2}+b_{2}$, so the second coefficient of $\left(
k+g_{1}\left(  \alpha\right)  \right)  \left(  k+g_{1}\left(  \alpha\right)
+1\right)  \cdots\left(  k+g_{2}\left(  \alpha\right)  \right)  $ is
subtracted from $c_{2}\left(  \alpha\right)  $. This coefficient is%
\begin{align*}
c_{2}^{\prime}\left(  \alpha\right)   &  :=\sum_{i=g_{1}\left(  \alpha\right)
}^{g_{2}\left(  \alpha\right)  }i=\frac{1}{2}\left(  g_{1}\left(
\alpha\right)  +g_{2}\left(  \alpha\right)  \right)  \left(  g_{2}\left(
\alpha\right)  -g_{1}\left(  \alpha\right)  +1\right) \\
&  =\frac{1}{2}\left(  5\alpha+1+\left\lfloor \frac{\alpha+1}{2}\right\rfloor
+\left\lfloor \frac{\alpha+1}{3}\right\rfloor \right)  \left(  \alpha
+\left\lfloor \frac{\alpha+1}{3}\right\rfloor -\left\lfloor \frac{\alpha+1}%
{2}\right\rfloor \right)  .
\end{align*}
The second coefficient of $\widetilde{p_{\alpha}}$ is $c_{2}\left(
\alpha\right)  -c_{2}^{\prime}\left(  \alpha\right)  $; the sequence of values
for $\alpha=1\ldots14$ is%
\[
\lbrack7,21,59,92,155,222,319,364,510,626,745,853,1068,1186].
\]

Denote the coefficient of $k^{4\alpha-4}$ in $\left(  \frac{2^{8\alpha+1}%
}{\left(  2\alpha-1\right)  !}\right)  ^{-1}p_{\alpha}\left(  k\right)  $ by
$c_{3}\left(  \alpha\right)  $ then from the calculated values ($\alpha
=1,\ldots,13$) we find for $\alpha\neq0\operatorname{mod}4$ that%
\[
c_{3}\left(  \alpha\right)  =11-\frac{389}{24}\alpha-\frac{333}{16}\alpha
^{2}+\frac{115}{48}\alpha^{3}+\frac{289}{8}\alpha^{4}.
\]

The third observation: when $\alpha$ is a half-integer then%
\[
F\left(  \alpha,k\right)  =\frac{p_{\alpha}\left(  k\right)  }{\left(
k+2\alpha+1\right)  _{\alpha+1/2}}G\left(  \alpha,k\right)  ,
\]
where $p_{\alpha}\left(  k\right)  $ is a polynomial of degree $5\alpha
-\frac{3}{2}$ with leading coefficient $\dfrac{2^{8\alpha+1}}{\left(
2\alpha-1\right)  !}$.
\section{Appendix II. A modification of the Provost-Legendre method using Gegenbauer polynomials}
We consider the problem of approximating a density function with given moments
using Jacobi polynomials for some choice of parameters. The technique uses a
construction of Provost \cite[sec. 4]{Provost} which is adapted for a specific aspect of the unknown probability density, namely, $\Pr\left\{  X>0\right\}  $.

Suppose the density $f\left(  x\right)  $ is supported on the interval
$\left[  a,b\right]  $ with given (i.e. computable) moments $\mu_{n}:=\int
_{a}^{b}x^{n}f\left(  x\right)  dx$, and $\left\{  p_{n}\left(  x\right)
\right\}  $ is a family of orthogonal polynomials  with weight function
$w\left(  x\right)  $ also on $\left[  a,b\right]  $; the structure constants
are%
\[
h_{n}:=\int_{a}^{b}p_{n}\left(  x\right)  ^{2}w\left(  x\right)
dx,~n=0,1,2,\ldots
\]
The aim is to (implicitly) determine the expansion%
\[
f\left(  x\right)  =\sum_{n=0}^{\infty}\lambda_{n}p_{n}\left(  x\right)
w\left(  x\right)  .
\]
and to apply it to the approximation of (where $a<0<b$)%
\[
\Pr\left\{  X>0\right\}  =\int_{0}^{b}f\left(  x\right)  dx=\sum_{n=0}%
^{\infty}\lambda_{n}\int_{0}^{b}p_{n}\left(  x\right)  w\left(  x\right)  dx.
\]
By orthogonality, for $m=0,1,2,\ldots$%
\begin{align*}
\int_{a}^{b}p_{m}\left(  x\right)  f\left(  x\right)  dx &  =\sum
_{n=0}^{\infty}\lambda_{n}\int_{a}^{b}p_{n}\left(  x\right)  p_{m}\left(
x\right)  w\left(  x\right)  dx\\
&  =\lambda_{m}h_{m}.
\end{align*}
To evaluate the left hand side compute the coefficients $\left\{  a_{ni}:0\leq
i\leq n\right\}  $ in the expansions%
\[
p_{n}\left(  x\right)  =\sum_{i=0}^{n}a_{ni}x^{i},
\]
when $\left\{  p_{n}\left(  x\right)  \right\}  $ are shifted Jacobi
polynomials (this requires extra computation since the shortest expansions are
in powers of $\left(  x-a\right)  $ or $\left(  b-x\right)  $); then%
\begin{align*}
\lambda_{m}h_{m} &  =\int_{a}^{b}\sum_{i=0}^{m}a_{mi}x^{i}f\left(  x\right)
dx=\sum_{i=0}^{m}a_{mi}\mu_{i},\\
\lambda_{m} &  =\frac{1}{h_{m}}\sum_{i=0}^{m}a_{mi}\mu_{i};
\end{align*}

The main problem is to approximate $\int_{\xi}^{b}f\left(  x\right)  dx$ for
some given $\xi$: so
\[
\int_{a}^{\xi}f\left(  x\right)  dx=\sum_{n=0}^{\infty}\lambda_{n}\int_{\xi
}^{b}p_{n}\left(  x\right)  w\left(  x\right)  dx.
\]
Compute%
\[
q_{n}:=\int_{\xi}^{b}p_{n}\left(  x\right)  w\left(  x\right)  dx,
\]
then%
\begin{align*}
\int_{\xi}^{b}f\left(  x\right)  dx &  =\sum_{n=0}^{\infty}\lambda_{n}%
q_{n}=\sum_{n=0}^{\infty}\frac{1}{h_{n}}\sum_{i=0}^{n}a_{ni}\mu_{i}q_{n}\\
&  =\sum_{i=0}^{\infty}\mu_{i}\sum_{n=i}^{\infty}\frac{q_{n}}{h_{n}}a_{ni},
\end{align*}
and now we observe that the sum over $n$ is the coefficient of $x^{i}$ in%
\[
\sum_{n=0}^{\infty}\frac{q_{n}}{h_{n}}p_{n}\left(  x\right)  .
\]
Truncate the infinite series to obtain an approximation.

\textbf{Jacobi polynomials}: We start with background information about
general parameters and then specialize to equal parameters. The family
$\left\{  P_{n}^{\left(  \alpha,\beta\right)  }\left(  t\right)  \right\}  $
is orthogonal for $\left(  1-t\right)  ^{\alpha}\left(  1+t\right)  ^{\beta}$;%
\begin{gather}
P_{n}^{\left(  \alpha,\beta\right)  }\left(  t\right)  =\frac{\left(
\alpha+1\right)  _{n}}{n!}~_{2}F_{1}\left(
\genfrac{}{}{0pt}{}{-n,n+\alpha+\beta+1}{\alpha+1}%
;\frac{1-t}{2}\right)  \nonumber\\
\frac{d}{dt}\left\{  \left(  1-t\right)  ^{\alpha+1}\left(  1+t\right)
^{\beta+1}P_{n-1}^{\left(  \alpha+1,\beta+1\right)  }\left(  t\right)
\right\}  =-2n\left(  1-t\right)  ^{\alpha}\left(  1+t\right)  ^{\beta}%
P_{n}^{\left(  \alpha,\beta\right)  }\left(  t\right)  \label{diffPnt}\\
h_{n}=2^{\alpha+\beta+1}B\left(  \alpha+1,\beta+1\right)  \frac{\left(
\alpha+1\right)  _{n}\left(  \beta+1\right)  _{n}\left(  \ \alpha
+\beta+n+1\right)  }{n!\left(  \alpha+\beta+2\right)  _{n}\left(
\ \alpha+\beta+2n+1\right)  }.\nonumber
\end{gather}
Equation (\ref{diffPnt}) is from \cite[18.9.16]{DLMF}. To shift to the
interval $\left[  a,b\right]  $ set%
\begin{align*}
x &  =\frac{1}{2}\left(  \left(  b-a\right)  t+a+b\right)  ,~t=\frac
{2x-a-b}{b-a}\\
w\left(  x\right)   &  =\left(  \frac{2}{b-a}\right)  ^{\alpha+\beta+1}\left(
b-x\right)  ^{\alpha}\left(  x-a\right)  ^{\beta},\\
p_{n}\left(  x\right)   &  =P_{n}^{\left(  \alpha,\beta\right)  }\left(
\frac{2x-a-b}{b-a}\right)  ,
\end{align*}
and the key quantities $q_{n}$ are found by%
\begin{align*}
\int_{\xi}^{b}p_{n}\left(  x\right)  w\left(  x\right)  dx &  =\left(
\frac{2}{b-a}\right)  ^{\alpha+\beta+1}\int_{\xi}^{b}P_{n}^{\left(
\alpha,\beta\right)  }\left(  \frac{2x-a-b}{b-a}\right)  \left(  b-x\right)
^{\alpha}\left(  x-a\right)  ^{\beta}dx\\
&  =\int_{\zeta}^{1}P_{n}^{\left(  \alpha,\beta\right)  }\left(  t\right)
\left(  1-t\right)  ^{\alpha}\left(  1+t\right)  ^{\beta}dt\\
&  =-\frac{1}{2n}\int_{\zeta}^{1}\frac{d}{dt}\left\{  \left(  1-t\right)
^{\alpha+1}\left(  1+t\right)  ^{\beta+1}P_{n-1}^{\left(  \alpha
+1,\beta+1\right)  }\left(  t\right)  \right\}  dt\\
&  =\frac{1}{2n}\left(  1-\zeta\right)  ^{\alpha+1}\left(  1+\zeta\right)
^{\beta+1}P_{n-1}^{\left(  \alpha+1,\beta+1\right)  }\left(  \zeta\right)
,n\geq1;\\
\zeta &  =\frac{2\xi-a-b}{b-a},
\end{align*}
and $q_{0}=\int_{\zeta}^{1}\left(  1-t\right)  ^{\alpha}\left(  1+t\right)
^{\beta}dt.$

In the case $a=-\frac{1}{16},b=\frac{1}{432}$, $\xi=0$ the transformations are%
\begin{align*}
t &  =\frac{216}{7}x+\frac{13}{14},\zeta=\frac{13}{14},\\
p_{n}\left(  x\right)   &  =P_{n}^{\left(  \alpha,\beta\right)  }\left(
\frac{216}{7}x+\frac{13}{14}\right)  .
\end{align*}
Thus the strategy is to choose appropriate parameters $\alpha,\beta$ (small
integer values appear to work well), then determine the coefficients of
$\left\{  x^{i}\right\}  $ in the truncated series%
\[
\sum_{n=0}^{\infty}\frac{q_{n}}{h_{n}}P_{n}^{\left(  \alpha,\beta\right)
}\left(  \frac{2x-a-b}{b-a}\right)  .
\]

\textbf{Computational details}:

Given $\left[  a,b\right]  $ with $a<0<b$ let $c_{0}:=-\dfrac{a+b}{b-a}$,
$c_{1}:=\dfrac{2}{b-a}$ and specialize to $\alpha=\beta=\lambda-\frac{1}%
{2}\geq0$, so that the weight is $\left(  1-t^{2}\right)  ^{\alpha}$ and the
Gegenbauer polynomials $P_{n}^{\lambda}$ form the orthogonal basis. We use the
normalized polynomials with $P_{n}^{\lambda}\left(  1\right)  =1$. (Note that
$P_{n}^{\lambda}\left(  t\right)  =\frac{n!}{\left(  \lambda+\frac{1}%
{2}\right)  _{n}}P_{n}^{\left(  \lambda-1/2,\lambda-1/2\right)  }\left(
t\right)  .)$ The recurrence is $P_{0}^{\lambda}\left(  t\right)
=1,P_{1}^{\lambda}\left(  t\right)  =t,$%
\[
P_{n+1}^{\lambda}\left(  t\right)  =\frac{2n+2\alpha+1}{n+2\alpha+1}%
tP_{n}^{\lambda}\left(  t\right)  -\frac{n}{n+2\alpha+1}P_{n-1}^{\lambda
}\left(  t\right)  ,n\geq1
\]
and%
\[
h_{n}=\frac{\Gamma\left(  \frac{1}{2}\right)  \Gamma\left(  \alpha+1\right)
}{\Gamma\left(  \alpha+\frac{3}{2}\right)  }\frac{n!\left(  2\alpha+1\right)
}{\left(  2\alpha+1\right)  _{n}\left(  2n+2\alpha+1\right)  }=h_{0}\eta_{n},
\]
where%
\[
\eta_{0}=1,\eta_{n}=\frac{n\left(  2n+2\alpha-1\right)  }{\left(
2n+2\alpha+1\right)  \left(  n+2\alpha\right)  }\eta_{n-1},n\geq1.
\]
(see \cite[Sect. 1.4.3]{DX}). In the recurrence replace $t$ by $c_{0}+y$,
where $y=c_{1}x$ (this takes the scaling factor out of the computations) Let%
\[
P_{n}^{\lambda}\left(  c_{0}+y\right)  =\sum_{j=0}^{n}B_{nj}y^{j},
\]
then (with the convention $B_{n,-1}=0$)%
\begin{align*}
B_{00} &  =1,~B_{1,0}=c_{0},~B_{1,1}=1,\\
B_{nj} &  =\frac{2n+2\alpha-1}{n+2\alpha}\left(  c_{0}B_{n-1,j}+B_{n-1,j-1}%
\right)  -\frac{n-1}{n+2\alpha}B_{n-2,j},~n\geq2,~0\leq j\leq n.
\end{align*}
Furthermore
\begin{gather*}
\frac{d}{dt}\left\{  \left(  1-t^{2}\right)  ^{\alpha+1}P_{n-1}^{\lambda
+1}\left(  t\right)  \right\}  =-2\left(  \alpha+1\right)  \left(
1-t^{2}\right)  ^{\alpha}P_{n}^{\lambda}\left(  t\right)  ,\\
q_{n}=\int_{c0}^{1}\left(  1-t^{2}\right)  ^{\alpha}P_{n}^{\lambda}\left(
t\right)  dt=\frac{1}{2\left(  \alpha+1\right)  }\left(  1-c_{0}^{2}\right)
^{\alpha+1}P_{n-1}^{\lambda+1}\left(  c_{0}\right)  ,n\geq1,\\
q_{0}=\int_{c_{0}}^{1}\left(  1-t^{2}\right)  ^{\alpha}dt,
\end{gather*}
and $P_{n-1}^{\lambda+1}\left(  c_{0}\right)  =g_{n}$ can be computed with the
recurrence%
\begin{align*}
g_{1}  & =1,g_{2}=c_{0},\\
g_{n}  & =\frac{2n+2\alpha-1}{n+2\alpha+1}c_{0}g_{n-1}-\frac{n-2}{n+2\alpha
+1}g_{n-2},
\end{align*}
thus $q_{1}=\frac{1}{2\left(  \alpha+1\right)  }\left(  1-c_{0}^{2}\right)
^{\alpha+1}$ and $q_{n}=g_{n}q_{1}$. Note: if $\alpha$ and $c_{0}$ are
rational then the quantities $\left\{  B_{nj}\right\}  $, $\left\{  \eta
_{n}\right\}  $ and $\left\{  g_{n}\right\}  $ can be computed in exact arithmetic.

Suppose the process is terminated at some $m$, then (approximate values)%
\begin{align*}
A_{0}  & =\frac{q_{0}}{h_{0}}+\frac{q_{1}}{h_{0}}\sum_{j=1}^{m}\frac{g_{j}%
}{\eta_{j}}B_{j,0}\\
A_{i}  & =c_{1}^{i}\frac{q_{1}}{h_{0}}\sum_{j=i}^{m}\frac{g_{j}}{\eta_{j}%
}B_{j,i},~1\leq i\leq m.
\end{align*}
Since polynomial interpolation tends to be not numerically well-conditioned (a
lot of cancellation) it is suggested to compute the quantities $\left\{
q_{j}\right\}  ,\left\{  B_{j,i}\right\}  $ to high precision, or even better,
in exact arithmetic for $\alpha=0,1,2,\ldots$.

{\bf{Conflict of Interests}} \newline
The authors declare that there is no conflict of interests regarding the publication of this paper.

\begin{acknowledgments}
PBS expresses appreciation to the Kavli Institute for Theoretical
Physics (KITP) for computational support in this research. 
\end{acknowledgments}

\bibliography{Hindawi2}

\begin{thebibliography}{22}
\expandafter\ifx\csname natexlab\endcsname\relax\def\natexlab#1{#1}\fi
\expandafter\ifx\csname bibnamefont\endcsname\relax
  \def\bibnamefont#1{#1}\fi
\expandafter\ifx\csname bibfnamefont\endcsname\relax
  \def\bibfnamefont#1{#1}\fi
\expandafter\ifx\csname citenamefont\endcsname\relax
  \def\citenamefont#1{#1}\fi
\expandafter\ifx\csname url\endcsname\relax
  \def\url#1{\texttt{#1}}\fi
\expandafter\ifx\csname urlprefix\endcsname\relax\def\urlprefix{URL }\fi
\providecommand{\bibinfo}[2]{#2}
\providecommand{\eprint}[2][]{\url{#2}}

\bibitem[{\citenamefont{{\.Z}yczkowski
  et~al.}(1998)\citenamefont{{\.Z}yczkowski, Horodecki, Sanpera, and
  Lewenstein}}]{ZHSL}
\bibinfo{author}{\bibfnamefont{K.}~\bibnamefont{{\.Z}yczkowski}},
  \bibinfo{author}{\bibfnamefont{P.}~\bibnamefont{Horodecki}},
  \bibinfo{author}{\bibfnamefont{A.}~\bibnamefont{Sanpera}}, \bibnamefont{and}
  \bibinfo{author}{\bibfnamefont{M.}~\bibnamefont{Lewenstein}},
  \bibinfo{journal}{Phys. Rev. A} \textbf{\bibinfo{volume}{58}},
  \bibinfo{pages}{883} (\bibinfo{year}{1998}).

\bibitem[{\citenamefont{Aubrun et~al.}(2014)\citenamefont{Aubrun, Szarek, and
  Ye}}]{aubrun2}
\bibinfo{author}{\bibfnamefont{G.}~\bibnamefont{Aubrun}},
  \bibinfo{author}{\bibfnamefont{S.~J.} \bibnamefont{Szarek}},
  \bibnamefont{and} \bibinfo{author}{\bibfnamefont{D.}~\bibnamefont{Ye}},
  \bibinfo{journal}{Commun. Pure Appl. Math.} \textbf{\bibinfo{volume}{LXVII}},
  \bibinfo{pages}{0129} (\bibinfo{year}{2014}).

\bibitem[{\citenamefont{Bhosale et~al.}(2012)\citenamefont{Bhosale, Tomsovic,
  and Lakshminarayan}}]{Bhosale}
\bibinfo{author}{\bibfnamefont{U.~T.} \bibnamefont{Bhosale}},
  \bibinfo{author}{\bibfnamefont{S.}~\bibnamefont{Tomsovic}}, \bibnamefont{and}
  \bibinfo{author}{\bibfnamefont{A.}~\bibnamefont{Lakshminarayan}},
  \bibinfo{journal}{Phys. Rev. A} \textbf{\bibinfo{volume}{85}},
  \bibinfo{pages}{062331} (\bibinfo{year}{2012}).

\bibitem[{\citenamefont{Singh et~al.}(2014)\citenamefont{Singh, Kunjwal, and
  Simon}}]{Simon}
\bibinfo{author}{\bibfnamefont{R.}~\bibnamefont{Singh}},
  \bibinfo{author}{\bibfnamefont{R.}~\bibnamefont{Kunjwal}}, \bibnamefont{and}
  \bibinfo{author}{\bibfnamefont{R.}~\bibnamefont{Simon}},
  \bibinfo{journal}{Phys. Rev. A} \textbf{\bibinfo{volume}{89}},
  \bibinfo{pages}{022308} (\bibinfo{year}{2014}).

\bibitem[{\citenamefont{{\.Z}yczkowski and Sommers}(2001)}]{Induced}
\bibinfo{author}{\bibfnamefont{K.}~\bibnamefont{{\.Z}yczkowski}}
  \bibnamefont{and} \bibinfo{author}{\bibfnamefont{H.-J.}
  \bibnamefont{Sommers}}, \bibinfo{journal}{J. Phys. A}
  \textbf{\bibinfo{volume}{A34}}, \bibinfo{pages}{7111} (\bibinfo{year}{2001}).

\bibitem[{\citenamefont{Bengtsson and {\.Z}yczkowski}(2006)}]{ingemarkarol}
\bibinfo{author}{\bibfnamefont{I.}~\bibnamefont{Bengtsson}} \bibnamefont{and}
  \bibinfo{author}{\bibfnamefont{K.}~\bibnamefont{{\.Z}yczkowski}},
  \emph{\bibinfo{title}{Geometry of Quantum States}}
  (\bibinfo{publisher}{Cambridge}, \bibinfo{address}{Cambridge},
  \bibinfo{year}{2006}).

\bibitem[{\citenamefont{Slater and Dunkl}(2012)}]{MomentBased}
\bibinfo{author}{\bibfnamefont{P.~B.} \bibnamefont{Slater}} \bibnamefont{and}
  \bibinfo{author}{\bibfnamefont{C.~F.} \bibnamefont{Dunkl}},
  \bibinfo{journal}{J. Phys. A} \textbf{\bibinfo{volume}{45}},
  \bibinfo{pages}{095305} (\bibinfo{year}{2012}).

\bibitem[{\citenamefont{Slater}(2013)}]{slaterJModPhys}
\bibinfo{author}{\bibfnamefont{P.~B.} \bibnamefont{Slater}},
  \bibinfo{journal}{J. Phys. A} \textbf{\bibinfo{volume}{46}},
  \bibinfo{pages}{445302} (\bibinfo{year}{2013}).

\bibitem[{\citenamefont{Fei and Joynt}()}]{FeiJoynt}
\bibinfo{author}{\bibfnamefont{J.}~\bibnamefont{Fei}} \bibnamefont{and}
  \bibinfo{author}{\bibfnamefont{R.}~\bibnamefont{Joynt}},
  \eprint{arXiv.1409:1993}.

\bibitem[{\citenamefont{Milz and Strunz}(2015)}]{milzstrunz}
\bibinfo{author}{\bibfnamefont{S.}~\bibnamefont{Milz}} \bibnamefont{and}
  \bibinfo{author}{\bibfnamefont{W.~T.} \bibnamefont{Strunz}},
  \bibinfo{journal}{J. Phys. A} \textbf{\bibinfo{volume}{48}},
  \bibinfo{pages}{035306} (\bibinfo{year}{2015}).

\bibitem[{\citenamefont{Khvedelidzea and Rogojina}(2013)}]{Dubna}
\bibinfo{author}{\bibfnamefont{A.}~\bibnamefont{Khvedelidzea}}
  \bibnamefont{and} \bibinfo{author}{\bibfnamefont{I.}~\bibnamefont{Rogojina}}
  (\bibinfo{year}{2013}), \eprint{Joint Institute for Nuclear Research, Dubna}.

\bibitem[{\citenamefont{Peres}(1996)}]{asher}
\bibinfo{author}{\bibfnamefont{A.}~\bibnamefont{Peres}},
  \bibinfo{journal}{Phys. Rev. Lett.} \textbf{\bibinfo{volume}{77}},
  \bibinfo{pages}{1413} (\bibinfo{year}{1996}).

\bibitem[{\citenamefont{{\.Z}yczkowski and Sommers}(2003)}]{szHS}
\bibinfo{author}{\bibfnamefont{K.}~\bibnamefont{{\.Z}yczkowski}}
  \bibnamefont{and} \bibinfo{author}{\bibfnamefont{H.-J.}
  \bibnamefont{Sommers}}, \bibinfo{journal}{J. Phys. A}
  \textbf{\bibinfo{volume}{36}}, \bibinfo{pages}{10115} (\bibinfo{year}{2003}).

\bibitem[{\citenamefont{Dumitriu and Edelman}(2002)}]{MatrixModels}
\bibinfo{author}{\bibfnamefont{I.}~\bibnamefont{Dumitriu}} \bibnamefont{and}
  \bibinfo{author}{\bibfnamefont{A.}~\bibnamefont{Edelman}},
  \bibinfo{journal}{J. Math. Phys.} \textbf{\bibinfo{volume}{43}},
  \bibinfo{pages}{5830} (\bibinfo{year}{2002}).

\bibitem[{\citenamefont{Slater}()}]{HyperDiff}
\bibinfo{author}{\bibfnamefont{P.~B.} \bibnamefont{Slater}},
  \eprint{arXiv:1504.04555}.

\bibitem[{\citenamefont{Horodecki et~al.}(1996)\citenamefont{Horodecki,
  Horodecki, and Horodecki}}]{michal}
\bibinfo{author}{\bibfnamefont{M.}~\bibnamefont{Horodecki}},
  \bibinfo{author}{\bibfnamefont{P.}~\bibnamefont{Horodecki}},
  \bibnamefont{and}
  \bibinfo{author}{\bibfnamefont{R.}~\bibnamefont{Horodecki}},
  \bibinfo{journal}{Phys. Lett. A} \textbf{\bibinfo{volume}{223}},
  \bibinfo{pages}{1} (\bibinfo{year}{1996}).

\bibitem[{\citenamefont{Provost}(2005)}]{Provost}
\bibinfo{author}{\bibfnamefont{S.~B.} \bibnamefont{Provost}},
  \bibinfo{journal}{Mathematica J.} \textbf{\bibinfo{volume}{9}},
  \bibinfo{pages}{727} (\bibinfo{year}{2005}).

\bibitem[{\citenamefont{Slater and Dunkl}(2015)}]{WholeHalf}
\bibinfo{author}{\bibfnamefont{P.~B.} \bibnamefont{Slater}} \bibnamefont{and}
  \bibinfo{author}{\bibfnamefont{C.~F.} \bibnamefont{Dunkl}},
  \bibinfo{journal}{J. Geom. Phys.} \textbf{\bibinfo{volume}{90}},
  \bibinfo{pages}{42} (\bibinfo{year}{2015}).

\bibitem[{\citenamefont{Dunkl and Slater}()}]{LatestCollaboration2}
\bibinfo{author}{\bibfnamefont{C.~F.} \bibnamefont{Dunkl}} \bibnamefont{and}
  \bibinfo{author}{\bibfnamefont{P.~B.} \bibnamefont{Slater}},
  \eprint{arXiv:1501.02289}.

\bibitem[{\citenamefont{Mendon\c{c}a et~al.}(2014)\citenamefont{Mendon\c{c}a,
  Marchiolli, and Galetti}}]{Xstates2}
\bibinfo{author}{\bibfnamefont{P.}~\bibnamefont{Mendon\c{c}a}},
  \bibinfo{author}{\bibfnamefont{M.~A.} \bibnamefont{Marchiolli}},
  \bibnamefont{and} \bibinfo{author}{\bibfnamefont{D.}~\bibnamefont{Galetti}},
  \bibinfo{journal}{Anns. Phys.} \textbf{\bibinfo{volume}{351}},
  \bibinfo{pages}{79} (\bibinfo{year}{2014}).

\bibitem[{\citenamefont{Olver et~al.}(2010)\citenamefont{Olver, Lozier,
  Boisvert, and Clark}}]{DLMF}
\bibinfo{author}{\bibfnamefont{F.}~\bibnamefont{Olver}},
  \bibinfo{author}{\bibfnamefont{D.}~\bibnamefont{Lozier}},
  \bibinfo{author}{\bibfnamefont{R.}~\bibnamefont{Boisvert}}, \bibnamefont{and}
  \bibinfo{author}{\bibfnamefont{C.}~\bibnamefont{Clark}},
  \emph{\bibinfo{title}{NIST Handbook of Mathematical Functions}}
  (\bibinfo{publisher}{Cambridge Univ. Press}, \bibinfo{address}{Cambridge},
  \bibinfo{year}{2010}).

\bibitem[{\citenamefont{Dunkl and Xu}(2014)}]{DX}
\bibinfo{author}{\bibfnamefont{C.}~\bibnamefont{Dunkl}} \bibnamefont{and}
  \bibinfo{author}{\bibfnamefont{Y.}~\bibnamefont{Xu}},
  \emph{\bibinfo{title}{Orthogonal Polynomials of Several Variables}}
  (\bibinfo{publisher}{Cambridge Univ. Press}, \bibinfo{address}{Cambridge},
  \bibinfo{year}{2014}).

\end{thebibliography}

\end{document}